# Branched polymers on branched polymers

*Bergfinnur Durhuus*[1]

Matematisk Institut, University of Copenhagen
Universitetsparken 5, 2100 Copenhagen Ø
Denmark

*Thordur Jonsson*[2][3]

Department of Theoretical Physics, University of Oxford
1 Keble Road, OX1 3NP
United Kingdom

**Abstract.** We study an ensemble of branched polymers which are embedded on other branched polymers. This is a toy model which allows us to study explicitly and in detail the reaction of a statistical system on an underlying geometrical structure, a problem of interest in the study of the interaction of matter and quantized gravity. We find a phase transition at which the embedded polymers begin to cover the basis polymers. At the transition point the susceptibility exponent $\gamma$ takes the value $3/4$ and the two-point function develops an anomalous dimension $1/2$.

[1] e-mail: durhuus@math.ku.dk
[2] e-mail: thjons@raunvis.hi.is
[3] Permanent address: University of Iceland, Dunhaga 3, 107 Reykjavik, Iceland

# 1 Introduction

In the past few years there has been a considerable progress in the study of Euclidean quantum gravity in two space–time dimensions by means of discretized random surfaces and matrix models, see e.g. [1] for a recent review. The case of pure gravity as well as gravity interacting with matter whose central charge $c$ is smaller than 1 is now rather well understood [11, 6, 7]. The case of $c > 1$ seems to be qualitatively different and require new methods.

The basic problem is to understand the reaction of matter fields on an underlying dynamical geometric structure. For the Ising model on a random surface the problem was solved explicitly in [10] and the susceptibility exponent $\gamma$ of the geometry was found to jump at the phase transition point of the Ising model. If one puts matter fields with $c \geq 1$ on a random surface the response of the surface to the matter fields changes drastically and there are strong reasons to believe that $c > 1$ theories correspond to branched polymers in some sense [4, 8, 9, 5], i.e. too much matter on a random surface drives it into a collapsed state.

Placing matter fields, Ising spins or gaussian scalars, on branched polymers has no effect on their statistical behaviour [4]. In this letter we introduce a new simple branched polymer model with a phase transition which can be analysed explicitly and shares some of the qualitative features of the Ising model transition. The idea is to embed a branched polymer onto another branched polymer. The embedded polymers renormalize the couplings of the underlying polymers and when the embedded polymers become critical they shift the critical exponents of the underlying polymers.

# 2 The model

In order to make the following calculations as simple and transparent as possible we take our underlying branched polymer model to be of the simplest type. We do not expect the results to be affected by more complicated polymers as long as the susceptibility exponent remains $\frac{1}{2}$. We assume that we have a branched polymer model where the vertices of polymers are either of order 1 or 3 and these vertices have equal weight. The polymers are assumed to have a cyclic ordering of the links that meet at any vertex, i.e. the polymers are embedded in a plane. We shall call these polymers the *big polymers*. The partition function is given by

$$Z(\mu) = \sum_B e^{-\mu |B|}, \tag{1}$$

where the sum runs over all rooted big polymers and $|B|$ is the number of links in $B$. By rooted polymer we mean that one vertex is singled out as the "root" and for convenience we assume that the root has order one but this is not essential. It is quite easy to calculate $Z(\mu)$, as well as the critical value of the coupling $\mu_c$ [2, 3]. One finds that

$$Z(\mu) = e^{-\mu}(1 + Z^2(\mu)) \tag{2}$$

and

$$e^{\mu_c} = 2, \quad Z(\mu_c) = 1. \tag{3}$$

As $\mu \to \mu_c$,

$$Z(\mu) = 1 - c\sqrt{\mu - \mu_c} + O(\mu - \mu_c), \tag{4}$$

where $c$ is a positive constant, corresponding to a susceptibility exponent $\gamma = \frac{1}{2}$ for the big polymer model.

We now define the *small polymer* model. A small polymer $b$ is by definition any rooted polymer that can be embedded in some big polymer so that no two vertices in the small polymer are mapped onto the same vertex in the big polymer and we also assume for convenience that the root of the small polymer is mapped onto the root of the big polymer but this assumption is not essential. We see that the vertices of the small polymers are restricted to have order 1, 2 or 3 in the present case since the vertices of the big polymer have order 1 or 3. The polymer on polymer model we wish to study is defined by the partition function

$$W(\mu, \beta) = \sum_B e^{-\mu|B|} \sum_{b \subset B} e^{-\beta|b|}. \tag{5}$$

In the sum over the small polymers $b$ it is convenient to adopt the convention of excluding the empty small polymer. By reasoning, similar to the one used to derive Eq. (2), one finds

$$W(\mu, \beta) = e^{-\mu-\beta}\left(1 + Z^2(\mu) + 2W(\mu, \beta)Z(\mu) + W^2(\mu, \beta)\right), \tag{6}$$

see Fig. 1. Solving this equation and using (2) to eliminate $\mu$ we find

$$W = -Z + e^\beta \frac{1 + Z^2}{2Z} - \frac{1}{2}\sqrt{((1 + Z^2)Z^{-1}e^\beta - 2Z)^2 - 4(1 + Z^2)}. \tag{7}$$

Here and in the sequel we drop the arguments of $Z$ and $W$ from the notation. It follows from Eq. (3) that $Z \leq 1$ for all values of $\mu \geq \mu_c$ so for large values of $\beta$ the singularity of $W(\mu, \beta)$, as $\mu$ is decreased with fixed $\beta$, is determined by the singularity of $Z$ alone and the argument of the square root in Eq. (7) does not vanish. In this case the critical value $\mu_0(\beta)$ of $\mu$ is independent of $\beta$ and equal to $\mu_c$, small polymers are suppressed and their presence does not affect the critical behaviour of the big polymers. If $\beta$ becomes so small that the argument of the square root vanishes as $\mu$ decreases to its critical value, then the critical exponent $\gamma$ of the susceptibility, defined by

$$\chi(\mu, \beta) = -\frac{\partial W}{\partial \mu} \sim (\mu - \mu_0(\beta))^{-\gamma}, \qquad (8)$$

is determined by the singularity of the square root. The smallest value of $\beta$ for which this can occur is given by $\beta = \beta_0 \equiv \log(1 + \sqrt{2})$. For this value of $\beta$ the quantity under the square root approaches zero like $\sqrt{\mu - \mu_c}$ as $\mu$ tends to its critical value, according to Eq. (4). It follows that the partition function $W$ approaches its limiting value like $(\mu - \mu_c)^{\frac{1}{4}}$ which implies that

$$\gamma = \frac{3}{4}. \qquad (9)$$

The reader may be surprised to see $\gamma > \frac{1}{2}$ in view of the "universal" bound $\gamma \leq \frac{1}{2}$. The explanation is that the embedded small polymers cannot be described by any local field on the big polymer.

For $\beta < \beta_0$ the partition function for the big polymers remains analytic as $\mu \to \mu_0(\beta) > \mu_c$. It follows that $\gamma = \frac{1}{2}$. In this region the small polymers almost cover the big polymer and have only the effect of renormalizing the cosmological constant $\mu$ for the big polymers. We can therefore regard the phase transition at $\beta_0$ as a "wetting transition" for the small polymers: for $\beta < \beta_0$ the small polymers wet the big polymers. We refer to the curve $(\mu_0(\beta), \beta)$ in the plane as the *critical line*.

It is natural to define a *wetting susceptibility* $\chi_w(\mu, \beta)$ by

$$\chi_w(\mu, \beta) = -\frac{\partial W}{\partial \beta} \qquad (10)$$

with an associated wetting susceptibility exponent $\gamma_w$ defined in analogy with the definition (8) of $\gamma$. Up to a finite normalization factor the wetting susceptibility is a measure of the average size of the small polymers. The above calculations show that $\chi_w$ does not diverge at the critical line for $\beta > \beta_0$, $\gamma_w = \frac{1}{4}$ at $\beta = \beta_0$ and $\gamma_w = \frac{1}{2}$ for $\beta < \beta_0$.

## 3 The two-point function

In order to calculate the two-point function we consider big polymers with two marked vertices, one of which is assumed to be the root where the small polymer is also rooted. Let $x$ be the number of links separating the two marked vartices, see

Fig. 2. We shall refer to $x$ as the distance between the marked vertices. Then we can express the two point function as

$$G(x) = e^{-\mu x} 2^{x-1} (1 + Z^2) \sum_{n=1}^{x-1} (Z+W)^n Z^{x-n-1} e^{-\beta n}$$
$$+ e^{-\mu x - \beta x} 2^{x-1} (Z+W)^{x-1} (Z^2 + 1 + W^2 + 2WZ), \qquad (11)$$

where $n$ is the distance along the shortest path between the marked vertices on the big polymer which is covered by the small polymer and the factor $2^{x-1}$ corresponds to the fact that the outgrowths from the path between the two marked vertices can be to either side. The last term on the right hand side of (11) corresponds to $n = x$. The sum in Eq. (11) is trivial and we find that

$$G(x) = e^{-\mu x} (2Z)^{x-1} (1 + Z^2) Q \qquad (12)$$
$$\times \left( \frac{1 - Q^{x-1}}{1 - Q} + Q^{x-2} e^{-\beta} \left( 1 + \frac{W^2 + 2WZ}{Z^2 + 1} \right) \right),$$

where

$$Q \equiv \left( 1 + \frac{W}{Z} \right) e^{-\beta}. \qquad (13)$$

For $\beta > \beta_0$ we have

$$Q < 1 \qquad (14)$$

for any $\mu \geq \mu_c$ and it follows that the mass $m$, governing the exponential decay of the two-point function, is given by

$$m(\mu, \beta) = \mu - \log(2Z)$$
$$= \log \frac{1 + Z^2}{2Z^2} \qquad (15)$$

in this region. We see that the mass vanishes according to

$$m(\mu, \beta) \sim (\mu - \mu_0(\beta))^{\frac{1}{2}} \qquad (16)$$

as $\mu \to \mu_0(\beta)$. The same result is obtained for $\beta = \beta_0$, the only difference being that now (14) becomes an equality at the critical point.

There is a curve $\mathcal{C}$ in the coupling constant plane, given by

$$Q = 1, \qquad (17)$$

where

$$G(x) \sim x e^{-mx} \qquad (18)$$

as $x \to \infty$, but for other values of the coupling constants, away from the critical line, the function $G(x)$ has pure exponential decay. The curve $\mathcal{C}$ can be thought of as separating the "wet" phase from the "dry" phase. For fixed $\beta < \beta_0$ let $\mu_1(\beta)$ be the value of $\mu$ where (17) holds. It is easy to check that $\mu_1(\beta)$ is uniquely determined. For $\mu \geq \mu_1(\beta)$ the mass is still given by (15) but for $\mu < \mu_1(\beta)$ we obtain

$$m(\mu, \beta) = \beta - \log\left(\frac{2Z^2 + 2WZ}{1 + Z^2}\right). \tag{19}$$

According to Eq. (7) the critical behaviour of the mass is again described by (16) so $\nu$, the the critical exponent of the mass, equals $\frac{1}{2}$ for all values of $\beta$.

Let us now consider the scaling limit. We fix $\beta$ and choose the coupling $\mu = \mu(a)$ to be a function of a scale parameter $a$ such that $\mu(a) \to \mu_0(\beta)$ as $a \to 0$ and

$$\lim_{a \to 0} \frac{m(\mu(a))}{a} = m^* > 0, \tag{20}$$

where $m^*$ is the "continuum" mass. For $\beta \neq \beta_0$ we are either in the wet phase or the dry phase for $\mu$ in a neighbourhood of the critical line so the scaling limit is given by

$$G^*(y) \equiv \lim_{a \to 0} G(y/a) = Ce^{-m^*y} \tag{21}$$

where $C$ is a constant, $y \in \mathbb{R}^+$ and $a$ tends to zero such that $y/a$ runs through positive integers.

For $\beta = \beta_0$ we see that

$$\frac{1 - Q^{a^{-1}}}{1 - Q} \sim \frac{1 - (1 - c_1\sqrt{a})^{a^{-1}}}{\sqrt{a}}$$
$$\sim \frac{1}{\sqrt{a}} \tag{22}$$

as $a \to 0$, where $c_1$ is a positive constant. It follows that the two-point function has an anomalous dimension $\eta = \frac{1}{2}$ for $\beta = \beta_0$ and we must multiply $G(y/a)$ in (21) by a prefactor (wave function renormalization) $\sqrt{a}$ in order for the scaling limit to exist and be nonzero. We see that Fisher's scaling relation, which in the present context takes the form

$$\gamma = (1 + \eta)\nu, \tag{23}$$

holds for all points on the critical line.

## 4 Discussion

In order to compare the model introduced in this letter and random surfaces with multiple Ising models (or Potts models) it is natural to consider the generalization of the present model to the case where we have $n$ independent small polymers interacting with one underlying big polymer. Unfortunately nothing new happens in this case as the reader can most easily convince himself of by studying the simple generalization of Eq. (7) to the case of two independent small polymers.

Another possibility would be to consider a branched polymer with a gas of embedded small polymers that could have their roots in arbitrary locations on the big polymer. We expect this model to exhibit a phase transition as small polymers condense. We also expect new features to arise if one considers a model of random walks on branched polymers, i.e. a model with a two-point function of the form

$$G(x) = \sum_B e^{-\mu|B|} \sum_{\omega \in \Omega(B,x)} e^{-\beta|\omega|}, \tag{24}$$

where $\Omega(B,x)$ is the collection of all random walks on $B$ that begin at the root and end at a distance $x$ from the root and $|\omega|$ is the number of steps in $\omega$. This amounts to studying diffusion on branched polymers which is an unsolved problem as far as we know.

The special case of self-avoiding walks on a branched polymer is easy since there is a unique self-avoiding walk on a branched polymer from the root to a given vertex. It follows that the evaluation of the two-point function corresponding to Eq. (24) is the same problem as calculating the average number of vertices at a given distance $x$ from the root on a branched polymer, denoted $\langle n(x) \rangle$. This problem was solved in [3] with the result

$$\langle n(x) \rangle = G(x) = Ce^{-mx}, \tag{25}$$

where $C$ is a constant and the mass $m$ goes to 0 at the critical point with a critical exponent $\frac{1}{2}$. Alternatively, we can of course think of a self avoiding walk on a branched polymer as a linear polymer embedded on a branched polymer so this model fits into the framework of small polymers embedded on big polymers but with different critical exponents.

We remark that Eq. (25) implies that $\langle n(x) \rangle$ is a *constant* at the critical point. In this sense the fractal dimension of branched polymers equals 1 in contradistinction to the usual intrinsic Hausdorff dimension $d_h$ which equals 2 [3] and is related to the critical exponent of the mass by the scaling relation $d_h \nu = 1$. The above considerations show that the intrinsic Hausdorff dimension is unrelated to the quantity $\langle n(x) \rangle$ in any branched polymer model since the latter quantity only depends on the power correction to the two point function at the critical point.

**Acknowledgement.** We are indebted to the Erwin Schrödinger Institute for hospitality and stimulating atmosphere.

Figure Caption.

**Fig. 1** A graphical illustration of Eq. (6). The shaded circles correspond to the full partition function for a small polymer embedded on a big polymer. The empty circles correspond to the partition function of the big polymer model alone. The first term corresponds to the polymer consisting of only two vertices and one link.

**Fig. 2** A graphical illustration of Eq. (11). The shaded circles correspond to the full partition function $W$ while the empty circles correspond to $Z$. The small polymer extends to a distance $n$ from the root. In the figure $x = 6$.

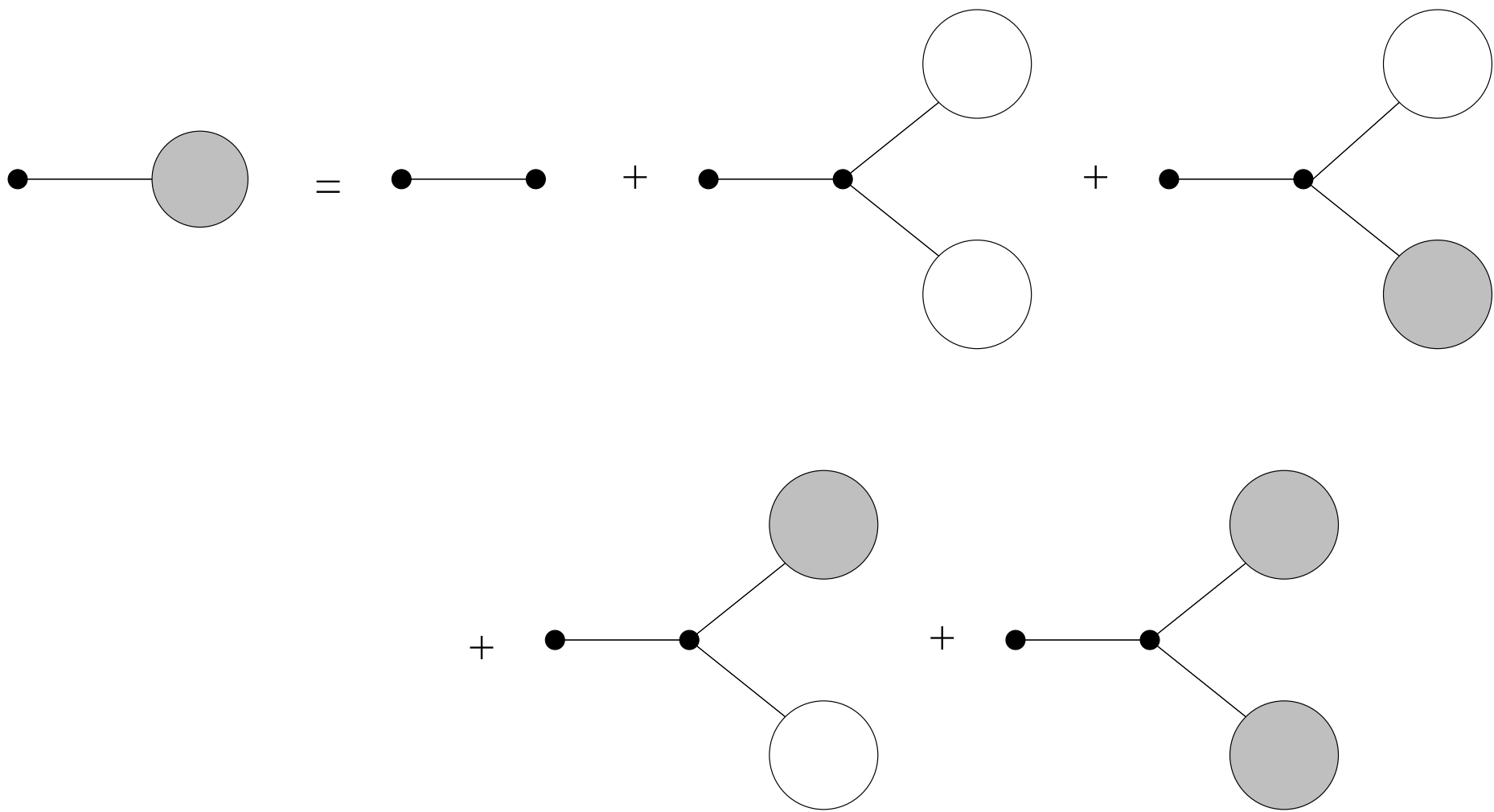

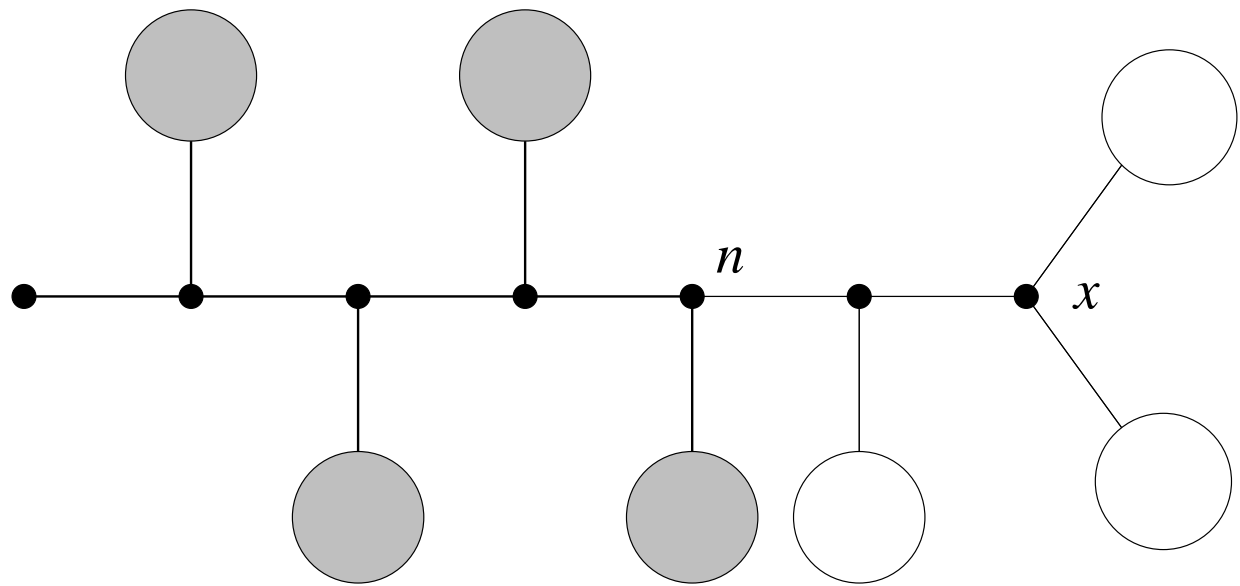